\documentclass[10pt,twocolumn]{elsart5p}
\usepackage{graphicx}
\usepackage{dcolumn}
\usepackage{amsmath}
\usepackage{amssymb}
\usepackage{bm}
\usepackage{natbib}

\begin{document}

\begin{frontmatter}

\title{Time-resolved and time-scale adaptive measures of spike train synchrony}

\author[ISC]{Thomas Kreuz\corauthref{cor}},
\corauth[cor]{Corresponding author.} \ead{thomas.kreuz@cnr.it}
\author[UPF]{Daniel Chicharro},
\author[SALK]{Martin Greschner},
\author[UPF]{Ralph G. Andrzejak},

\address[ISC]{Institute for Complex Systems, CNR, Sesto Fiorentino, Italy}
\address[UPF]{Department of Information and Communication Technologies, Universitat Pompeu Fabra, Barcelona, Spain}
\address[SALK]{The Salk Institute for Biological Studies, San Diego, California, USA}


\date{\today}

\begin{abstract}
    A wide variety of approaches to estimate the degree of synchrony between two or more spike trains have been proposed. One of the most recent methods is the ISI-distance which extracts information from the interspike intervals (ISIs) by evaluating the ratio of the instantaneous firing rates. In contrast to most previously proposed measures it is parameter free and time-scale independent. However, it is not well suited to track changes in synchrony that are based on spike coincidences. Here we propose the SPIKE-distance, a complementary measure which is sensitive to spike coincidences but still shares the fundamental advantages of the ISI-distance. In particular, it is easy to visualize in a time-resolved manner and can be extended to a method that is also applicable to larger sets of spike trains. We show the merit of the SPIKE-distance using both simulated and real data.
\end{abstract}


\end{frontmatter}

\newcommand{\abb}{\small\sf}

%
%
\section{\label{s:Intro} Introduction}

One of the key challenges of neuroscience is to understand the neuronal code, the way information is represented in the neurons' spike trains. Determining whether different stimuli can be distinguished from the spike train responses they trigger requires a notion of similarity of spike trains. This notion is typically expressed as a distance between spike trains. Reflecting the variety of neuronal coding hypotheses, several spike train distances with sensitivities to different features have been proposed \citep{Victor97, Kreuz07c}.
Some of these distances follow the classical view that firing rates play a central role in neural coding \citep{Barlow72}, and are thus based on the spike count \citep{Victor96}. However, in the last decades the fundamental relevance of temporal structure has been established \citep{Theunissen95}, and accordingly many distances focus on the timing of spikes \citep{Victor96, VanRossum01, S_Schreiber03}. Special cases of temporal coding are addressed by methods that deal with, e.g., the detection of coincidences \citep{Gruen99} and the identification of temporal spiking patterns \citep{Abeles88, Victor97}.

One of the most widely used spike train distances is the metric based on spike times introduced in \citet{Victor96} which evaluates the cost needed to transform one spike train into the other using only certain elementary steps. This method involves one parameter that sets the time-scale. In contrast, a more recent approach, the ISI-distance, is parameter free and time-scale adaptive \citep{Kreuz07c}. The ISI-distance is also easy to visualize in a time-resolved manner. On spike trains extracted from a simulated Hindemarsh-Rose network it could reproduce a known clustering equally well as the best time-scale-optimized measure \citep{Kreuz07c}. Recently, two multiple spike train extensions of the ISI-distance have been proposed, the local pairwise average and a multivariate measure based on the coefficient of variation \citep{Kreuz09a}.

The ISI-distance and its extensions are based on the relative length of simultaneous interspike intervals (ISI) and are thus well-designed to quantify similarities in the neurons' firing rate profiles. However, they are not optimally suited to track the type of synchrony which is mediated by spike timing and in particular by changes in the fraction of coincident spikes. This specific kind of sensitivity is not only of theoretical importance but also of high practical relevance, since coincidences of spikes have been proven to be of high prevalence in many different neuronal circuits. Examples include olfactory bulb \citep{Wilson06}, hippocampus \citep{Best01}, somatosensory cortex \citep{Romo03}, auditory cortex \citep{Schreiner07}, visual cortex \citep{Usrey99, Priebe08}, and retina \citep{Meister99, Field07, Shlens08}.

Therefore, motivated by both the importance of temporal coding and the ubiquity of coincident neuronal firing, we here propose the SPIKE-distance, a measure which uniquely combines the advantages of the ISI-distance - such as being parameter free, time-scale adaptive, and time-resolved - with a specific focus on spike timing. Given two spike trains, this focus is obtained by averaging for each time instant the absolute differences between the previous spike times and the following spike times and normalizing by the mean length of the interspike intervals. Furthermore, also for the SPIKE-distance we propose two extensions to the case of multiple spike trains, either as a local pairwise average or as a multivariate measure.

We use several sets of specifically designed and simulated spike trains in order to illustrate the properties of the SPIKE-distance and compare its performance to the one of previously published methods such as the ISI-distance, the Victor-Purpura distance and the correlation coefficient. In a first application to real data we employ the spiking activity of a population of retinal ganglion cells \citep{Meister99, Field07} as an ideal testing ground for the different measures. In particular, we evaluate their capability to reproduce the gradual decrease of synchrony between two retinal ganglion cells with the distance between their receptive fields \citep{Meister95, Shlens06}. As a second application to real data we employ the spike train distances to discriminate single-unit responses to taste stimuli recorded in the nucleus of the solitary tract (NTS) in rats (\citeauthor{DiLorenzo03}, \citeyear{DiLorenzo03}; data available online at http://neurodatabase.org/, \citeauthor{Gardner04}, \citeyear{Gardner04}). In this case the performance of the various measures is quantified by the normalized mutual information of the confusion matrix \citep{Abramson63, Victor96}.

The remainder of the paper is organized as follows: In Section \ref{ss:Methods-1-Spike-Distances} we describe the new SPIKE-distance and its extensions. Subsequently, we use constructed examples to stress the different sensitivities of the ISI- and the SPIKE-distance (Section \ref{ss:Methods-2-Sensitivity}), and contrast their local behavior to the global properties of the Victor-Purpura distance (Section \ref{ss:Methods-3-Local-behaviour}). We evaluate their capability to distinguish different levels of spike train synchrony (Section \ref{ss:Results-1-Capability}) and compare their performance on two real datasets (Section \ref{ss:Results-2-Application-to-Real-Data}). Finally, we provide some details on the derivation of the SPIKE-distance \ref{App-s:SPIKE-distance}, and describe the previously introduced measures against which we compare (Appendix \ref{App-s:Existing-Methods}) as well as the data that we use (Appendix \ref{App-s:Data}). In Appendix \ref{App-s:HR-Comparison} we evaluate the performance of the new SPIKE-distance in the setups used in \citeauthor{Kreuz07c} (\citeyear{Kreuz07c}, \citeyear{Kreuz09a}).

%
%

\section{\label{s:Methods} Methods}

In the following we will introduce the bivariate SPIKE-distance and its extensions. Like the ISI-distance (cf. Appendix \ref{App-ss:ISI-Distances}) all these measures are based on instantaneous values, i.e., from the sequences of spike times we create time profiles with one value for each sampling point. The distances are then defined as the temporal average of the respective time profile.

\subsection{\label{ss:Methods-1-Spike-Distances} The bivariate SPIKE-distance and its extensions}

We denote with $\{t_i^{(n)}\} = {t_1^{(n)},...,t_{M_n}^{(n)}}$ the spike times and with $M_n$ the number of spikes for neuron $n$ with $n = 1,...,N$.

\subsubsection{\label{sss:Bivariate-Spike-Distance} Bivariate SPIKE-distance}

\begin{figure}
    \includegraphics[width=85mm]{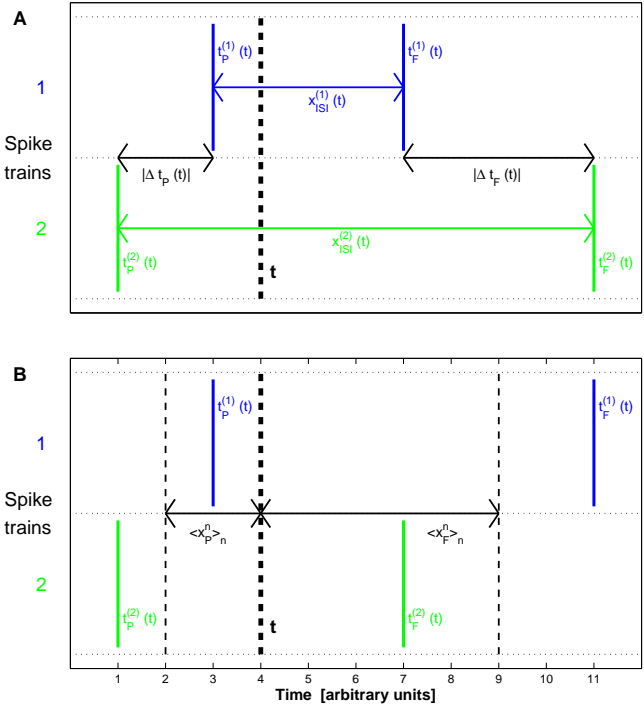}
    \caption{\abb\label{fig:Spike-Distance-Example} (color online) Illustration of the bivariate SPIKE-distance.   A. Time instant $t$ and its relation to the previous spikes $t_{\mathrm {P}}^{(n)}$ and the following spikes $t_{\mathrm {F}}^{(n)}$ as well as their differences. In this example the interspike interval of the first spike train is included in the interspike interval of the second spike train. The spikes of the two spike trains as well as their ISIs are shown in blue and green, respectively. The time instant under consideration is marked by a vertical dashed line, whereas horizontal solid arrows indicate the absolute differences $| \Delta t_{\mathrm {P}} (t) |$ and $| \Delta t_{\mathrm {F}} (t) |$ between the previous and the next spikes, respectively. B. Illustration of the weighting of the two contributions by the distance to the previous and the following spikes. Here the second spike of the second spike train comes first. The average distances to the previous and to the following spikes are represented by horizontal black lines. Note that for both examples we get the same value $S (t) = 0.367$ (Eq. \ref{eq:Bi-Spike-Diss2}), since we just switched the following spikes which effected the individual intervals but not their sum.}
\end{figure}

For each neuron $n = 1, 2$ we assign to each time instant the time of the previous spike
\begin{equation} \label{eq:Prev-Spike}
    t_{\mathrm {P}}^{(n)} (t) = \max(t_i^{(n)} | t_i^{(n)} \leq t)  \quad t_1^{(n)} \leq t \leq t_{M_n}^{(n)},
\end{equation}
and the time of the following spike
\begin{equation} \label{eq:Foll-Spike}
    t_{\mathrm {F}}^{(n)} (t) = \min(t_i^{(n)} | t_i^{(n)} > t)  \quad t_1^{(n)} \leq t \leq t_{M_n}^{(n)},
\end{equation}
as well as the interspike interval
\begin{equation} \label{eq:ISI}
    x_{\mathrm {ISI}}^{(n)} (t) = t_{\mathrm {F}}^{(n)} (t) - t_{\mathrm {P}}^{(n)} (t).
\end{equation}
We denote the instantaneous differences of previous and following spike times as
\begin{equation} \label{eq:Prev-Diff}
    \Delta t_{\mathrm {P}} (t) = t_{\mathrm {P}}^{(1)} (t) - t_{\mathrm {P}}^{(2)} (t)
\end{equation}
and
\begin{equation} \label{eq:Foll-Diff}
    \Delta t_{\mathrm {F}} (t) = t_{\mathrm {F}}^{(1)} (t) - t_{\mathrm {F}}^{(2)} (t),
\end{equation}
respectively. Note that by definition the intervals $\Delta t_{\mathrm {P}} (t)$ and $\Delta t_{\mathrm {F}} (t)$ do not cover the time instant $t$ (see Fig. \ref{fig:Spike-Distance-Example}A).

An instantaneous spike time based measure of spike train distance is given by the regular mean of the two absolute differences of previous and following spike times. This function is piecewise constant taking just one value within each interval of the pooled spike train. In order to be more local in time, we weight the two absolute differences $| \Delta t_{\mathrm {P}} (t) |$ and $| \Delta t_{\mathrm {F}} (t) |$ such that the difference of the spikes that are closer dominate (Fig. \ref{fig:Spike-Distance-Example}B). To this aim we denote with
\begin{equation} \label{eq:Prev-Spike-Dist}
     x_{\mathrm {P}}^{(n)} (t) = t - t_{\mathrm {P}}^{(n)} (t)
\end{equation}
and
\begin{equation} \label{eq:Foll-Spike-Dist}
     x_{\mathrm {F}}^{(n)} (t) = t_{\mathrm {F}}^{(n)} (t) - t
\end{equation}
the intervals to the previous and the following spikes for each neuron $n = 1, 2$. The inverse values of their respective averages over both neurons $\langle x_{\mathrm {P}}^{(n)} (t) \rangle_n$ and $\langle x_{\mathrm {F}}^{(n)} (t) \rangle_n$ serve as weights.

Since the relevance of a certain spike time difference depends on the local rate (e.g., a small shift within a burst is of relatively higher importance than the same small shift between two isolated spikes), we normalize the average spike time difference by the mean interspike interval (here $N = 2$)
\begin{equation} \label{eq:Mean-ISI}
    \langle x_{\mathrm {ISI}}^{(n)} (t) \rangle_n = \frac{1}{N} \sum_{n=1}^N x_{\mathrm {ISI}}^{(n)} (t).
\end{equation}
This way we also obtain time-scale invariance, i.e., the normalized spike time difference is the same for stretched or compressed spike trains. Using these quantities for locally weighted averaging and normalization we obtain the instantaneous spike time difference as (for a complete and more detailed derivation cf. Appendix \ref{App-s:SPIKE-distance}):
\begin{equation} \label{eq:Bi-Spike-Diss2}
    S (t) = \frac{| \Delta t_{\mathrm {P}} (t) | \langle x_{\mathrm {F}}^{(n)} (t) \rangle_n + | \Delta t_{\mathrm {F}} (t) | \langle x_{\mathrm {P}}^{(n)} (t) \rangle_n}{\langle x_{\mathrm {ISI}}^{(n)} (t) \rangle_n^2}.
\end{equation}
Finally, integrating over time yields the bivariate SPIKE-distance $D_{\mathrm {S}}$:
\begin{equation} \label{eq:Bi-Spike-Distances}
    D_{\mathrm {S}} = \frac{1}{T} \int_{t=0}^T S (t) dt.
\end{equation}
The bivariate SPIKE-distance is bounded in the interval $[0, 1]$. The value $0$ is only obtained for perfectly identical spike trains.

\subsubsection{\label{sss:Averaged-Spike-Distance} Averaged bivariate SPIKE-distance}

The averaged bivariate SPIKE-distance for a larger number of spike trains is the bivariate SPIKE-distance averaged over all pairs of neurons. The same kind of time-resolved visualization as in the bivariate case is possible, because the two averages over time and over pairs of neurons commute. We thus can first calculate the instantaneous average $S^{\mathrm {a}} (t)$ over all pairwise instantaneous spike time differences $S^{mn} (t)$ (Eq. \ref{eq:Bi-Spike-Diss2})
\begin{equation} \label{eq:Instantaneous-Spike-Average}
    S^{\mathrm {a}} (t) = \frac{1}{N(N-1)/2}\sum_{n=1}^N \sum_{m=n+1}^N S^{mn} (t)
\end{equation}
and then average over time
\begin{equation} \label{eq:Av-Bi-spike distance}
    D_{\mathrm {S}}^{\mathrm {a}} = \frac{1}{T} \int_{t=0}^T S^{\mathrm {a}} (t) dt.
\end{equation}
Like the original SPIKE-distance, the averaged bivariate SPIKE-distance is restricted to the interval $[0, 1]$.

\subsubsection{\label{sss:Multivariate-Spike-Distance} Multivariate SPIKE-distance}

Calculating the average over pairs of spike trains can become very time-consuming as soon as the number of neurons $N$ increases, since the computational cost scales with $N^2$. In such cases a multivariate approach which scales only with $N$ is computationally preferable.

For each time instant we take the standard deviations over all previous and all following spike times and divide their mean by the mean instantaneous ISI. Similar to the bivariate case (cf. Section \ref{sss:Bivariate-Spike-Distance} and Appendix \ref{App-s:SPIKE-distance}), in order to be more local in time we weight the two terms such that the standard deviation of the spike times that are closer dominates. This way we obtain the instantaneous spike time standard deviation:
\begin{equation} \label{eq:Multi-Spike-Diss2}
    S^{\mathrm {m}} (t) = \frac{\sigma [t_{\mathrm {P}}^{(n)} (t)]_n \langle x_{\mathrm {F}}^{(n)} (t) \rangle_n + \sigma [t_{\mathrm {F}}^{(n)} (t)]_n \langle x_{\mathrm {P}}^{(n)} (t) \rangle_n}{\langle x_{\mathrm {ISI}}^{(n)} (t) \rangle_n^2}.
\end{equation}
Finally, the multivariate SPIKE-distance $D_{\mathrm {S}}^{\mathrm {m}}$ is obtained by integrating over time:
\begin{equation} \label{eq:Multi-Spike-Distances}
    D_{\mathrm {S}}^{\mathrm {m}} = \frac{1}{T} \int_{t=0}^T S^{\mathrm {m}} (t) dt.
\end{equation}
The multivariate SPIKE-distance has $0$ as a lower bound but does not have an upper bound and in particular can attain values larger than $1$.

\subsubsection{\label{sss:Averaging and calculation} Practical considerations}

Since $\langle x_{\mathrm {P}}^{(n)} (t) \rangle_n$ and $\langle x_{\mathrm {F}}^{(n)} (t) \rangle_n$ depend explicitly on the time instant $t$, all three types of instantaneous SPIKE-measures $S (t)$, $S^{\mathrm {a}} (t)$, and $S^{\mathrm {m}} (t)$ in Eqs. \ref{eq:Bi-Spike-Diss2}, \ref{eq:Instantaneous-Spike-Average}, \ref{eq:Multi-Spike-Diss2} are piecewise linear rather than piecewise constant. Thus a new value has to be calculated for each sampling point and not just once per each interval in the pooled spike train. However, this is only necessary when the localized visualization is desired. In case the distance value itself is sufficient, the short computation time can be even further decreased by representing each interval by the value of its center and weighting it by its length. This actually gives the correct result, the time-resolved calculation is a very good approximation for sufficiently small sampling intervals $dt$.

An alternative variant to the time-weighted average used in Eqs. \ref{eq:Bi-Spike-Distances}, \ref{eq:Av-Bi-spike distance}, and \ref{eq:Multi-Spike-Distances} is spike-weighted averaging where the SPIKE-distances are evaluated only after every new spike. This way each interval of the pooled spike trains contributes the same and not according to its length. For the sake of brevity in this study we restrict ourselves to presenting results of the time-weighted average only, although the spike-weighted variant occasionally (for example for the data analyzed in Section \ref{sss:Results-3-Application-to-StimDis}) exhibits slightly better results.

A technical detail concerns the ambiguous definition of the very first and the very last interspike interval as well as the initial distance to the previous spike and the final distance to the following spike. This issue can be resolved by placing for each spike train auxiliary leading spikes in the beginning of the recording at time $t = 0$ and auxiliary trailing spikes at the end of the recording at time $t = T$. This leads to a systematic underestimation of the distance at the edges. While this limits the applicability of these methods to very sparse spike trains, the effect fades quickly with increasing numbers of spikes. A similar procedure has been used in the metric $D^{\mathrm {interval:fix}}$ in \citet{Victor97}.

More details on the implementation as well as the Matlab source code for calculating and visualizing both the SPIKE- and the ISI-distances can be found under www.fi.isc.cnr.it/users/thomas.kreuz/sourcecode.html.

\subsection{\label{ss:Methods-2-Sensitivity} Motivation: Sensitivity to spike coincidences and the spike train of origin}

\begin{figure}
    \includegraphics[width=85mm]{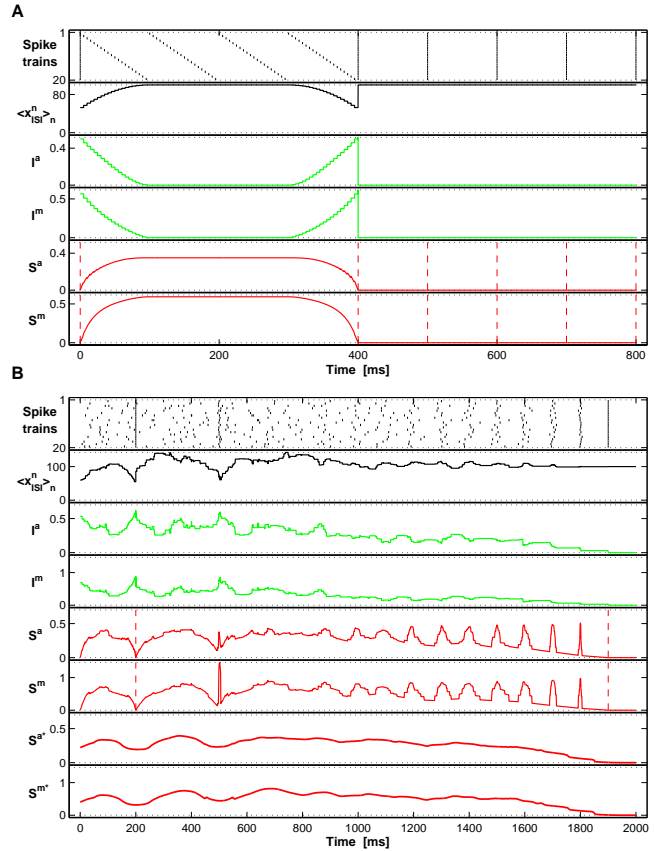}
    \caption{\abb\label{fig:Constructed-Example} (color online) Instantaneous ISI- and SPIKE-measures for constructed spike trains. On top we depict the spike trains and the mean instantaneous ISIs according to Eq. \ref{eq:ISI}. Below that we show the four instantaneous measures (from top to bottom): the averaged bivariate ISI-ratio $I^{\mathrm {a}} (t)$ (Eq. \ref{eq:Instantaneous-ISI-Average}), the multivariate coefficient of variation of the interspike intervals $I^{\mathrm {m}} (t)$ (Eq. \ref{eq:ISI-CV}), the spike time difference $S^{\mathrm {a}} (t)$ (Eq. \ref{eq:Bi-Spike-Diss2}), and finally the spike time standard deviation $S^{\mathrm {m}} (t)$ (Eq. \ref{eq:Multi-Spike-Diss2}). For the latter dashed vertical lines mark the occurrence of perfectly synchronous events.   A. In the first interval all spike trains are regular, but with constant phase lags, then all spikes are regular with zero phase lag, i.e., identical.   B. Spiking events with increasing jitter plus one noisy and one perfectly synchronous spiking event towards the end. For the instantaneous SPIKE-measures we also show the moving averages (of order $50$ sampling intervals) at the bottom of this subplot.}
\end{figure}

The main difference between the SPIKE-based methods and the previously proposed ISI-based methods (cf. Appendix \ref{App-ss:ISI-Distances}) is their different sensitivity to spike coincidences. The multiple spike trains shown in Fig. \ref{fig:Constructed-Example} were specifically constructed to illustrate this. In the first half ($< 400$ ms) of Fig. \ref{fig:Constructed-Example}a all $20$ spike trains are regular but with constant phase lags, while in the second half ($400 - 800$ ms) all spike trains are identical. Neither the averaged bivariate ISI-distance $D_{\mathrm {I}}^{\mathrm {a}}$ nor the multivariate ISI-distance $D_{\mathrm {I}}^{\mathrm {m}}$ distinguish between the intervals $100 - 300$ ms (the first interval without the transients) and $400 - 800$ ms, since the instantaneous frequencies are the same thus yielding a value of zero in both cases. The SPIKE-distances $D_{\mathrm {S}}^{\mathrm {a}}$ and $D_{\mathrm {S}}^{\mathrm {m}}$ on the other hand do distinguish, and higher values are obtained for the first interval in which spikes are maximally dispersed, whereas the coincident spikes in the second interval yield zero values indicating perfect synchrony.

In the example of Fig. \ref{fig:Constructed-Example}b we start with spike trains that are uncorrelated and basically random. At $200$ ms and at $500$ ms two events are inserted, the first one perfectly synchronous and the second one with a certain amount of jitter. Subsequently, the spike trains become more and more regular which is achieved by creating events every $100$ ms with decreasing amounts of jitter and closing with a perfectly synchronous spiking event at $1900$ ms.

The perfectly synchronous spiking event embedded within the noise ($200$ ms) is recognized only by the two instantaneous SPIKE-measures $S^{\mathrm {a}}$ and $S^{\mathrm {m}}$ which decrease to zero. On the other hand, the instantaneous ISI-measures even show peaks which are caused by the normalization. Similar peaks occur at the second unreliable event ($500$ ms). The SPIKE-measures on the other hand mark this event with a Mexican hat shaped event kernel~\footnote{Here the expression "Mexican hat" is not meant to strictly describe the difference between two Gaussians, but is rather used in a general sense to describe the succession "plateau - small drop - high peak - small drop -plateau".}. The reason for the Mexican hat shape of these event kernels is intuitive: Right before the spiking event all differences between the following spike times are small as are all differences between the previous spikes right after the event. This affects respectively the first and the second half of the numerator entering the instantaneous dissimilarities (cf. Eq. \ref{eq:Bi-Spike-Diss2}). Accordingly, values are decreased with respect to the mean level. Furthermore, within the spiking event there is a small interval for which some of the spikes are still following spikes while others are already previous spikes. Thus for this small interval the respective differences between previous and following spike times are very large and so is the value of the instantaneous SPIKE-measures.

Subsequently, when the noise is gradually decreased, all measures, starting from a plateau with moderate fluctuations reflecting the initial inherent randomness, exhibit a consistent decrease. However, this drop occurs in a different manner for the instantaneous ISI- and SPIKE-measures. While the ISI-measures decrease rather gradually with only a few smaller elevations, the SPIKE-measures mark the events with a series of peaks. These peaks are no longer Mexican hat shaped because between the events there is no noisy background level to decline to. With the increasing reliability of the spike events the peaks are getting more and more prominent. Consequently, a perfectly synchronous event like the one at $1900$ ms could be represented as a peak with infinitely small width and maximum amplitude. For such events the SPIKE-measures $S^{\mathrm {a}}$ and $S^{\mathrm {m}}$ yield the value zero (as expected for perfectly synchronous spike trains). The same value is also obtained for the intervals between two perfectly synchronous events. In order to distinguish these two cases we mark the occurrences of the perfect events by adding vertical dashed lines to the temporal profiles (see second half of Fig. \ref{fig:Constructed-Example}A).

The instantaneous SPIKE-measures are sensitive to spike coincidences and assign specific signatures to their time of occurrence. However, in case the focus of attention lies on the long-term changes in spike train synchrony, an appropriate moving average eliminates these short time-scale signatures and a gradual decrease similar to the one of the ISI-measures is obtained (cf. the two lowest subplots of Fig. \ref{fig:Constructed-Example}B).

\begin{figure}
    \includegraphics[width=85mm]{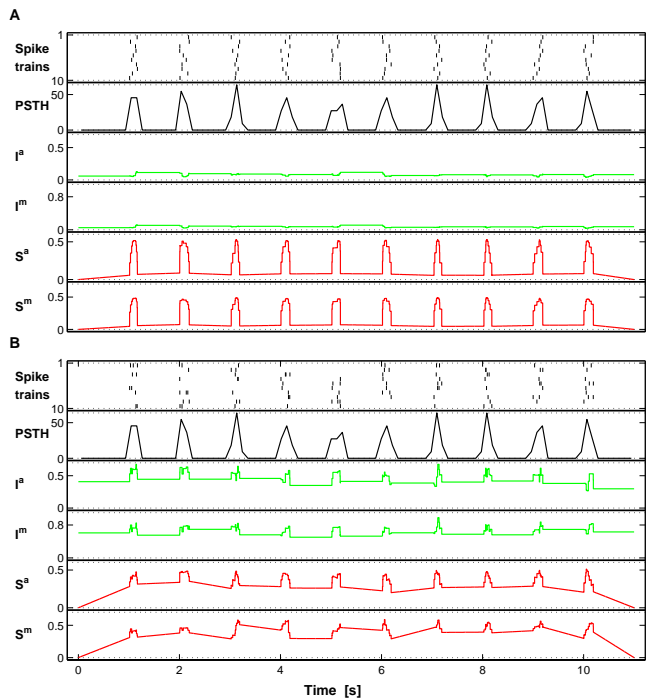}
    \caption{\abb\label{fig:SPIKE-PSTH} Two sets of spike trains demonstrating the sensitivity of the instantaneous ISI- and SPIKE-measures to the spike train of origin. By construction the pooled spike train of the two sets is identical consisting of $10$ regularly distributed bursts. Only the distribution of the spikes among the individual spike trains differs.   A. High reliability: Spikes from each burst are equally distributed among the spike trains.   B. Irregular behaviour: All spikes are randomly assigned to the individual spike trains.    Whereas the PSTH is by construction exactly identical in both cases, the ISI- and the SPIKE-distance can distinguish these cases resulting in low distances in the first case and in higher distances in the second case.}
\end{figure}

The SPIKE-distances are not only sensitive to the timing of the individual spikes, but also to their spike train of origin (Fig. \ref{fig:SPIKE-PSTH}). To show this, we start with one spike train consisting of $10$ equally distributed bursts of $10$ spikes each. From this spike train we construct two sets of $N = 10$ spike trains with $M_n = 10$ spikes each. This is achieved by assigning the individual spikes to their spike train of origin which is the inverse operation to pooling spikes from several spike trains. In the first set the spikes are distributed such that we have $10$ unreliable events with one spike in each spike train (Fig. \ref{fig:SPIKE-PSTH}A), while in the second set the spikes are assigned to random spike trains keeping only the number of $M_n = 10$ spikes per spike train constant (Fig. \ref{fig:SPIKE-PSTH}B). As can be seen in the four lower traces, both the ISI- and the SPIKE-distances can clearly distinguish these qualitatively different behaviors. This is in contrast to the Peri-Stimulus Time Histogram (PSTH) which, like all other measures of spike train synchrony based on the PSTH or on the pooled spike train in general, is invariant to shuffling spikes among the spike trains.

\subsection{\label{ss:Methods-3-Local-behaviour} Properties: Local behavior}

\begin{figure}
    \includegraphics[width=85mm]{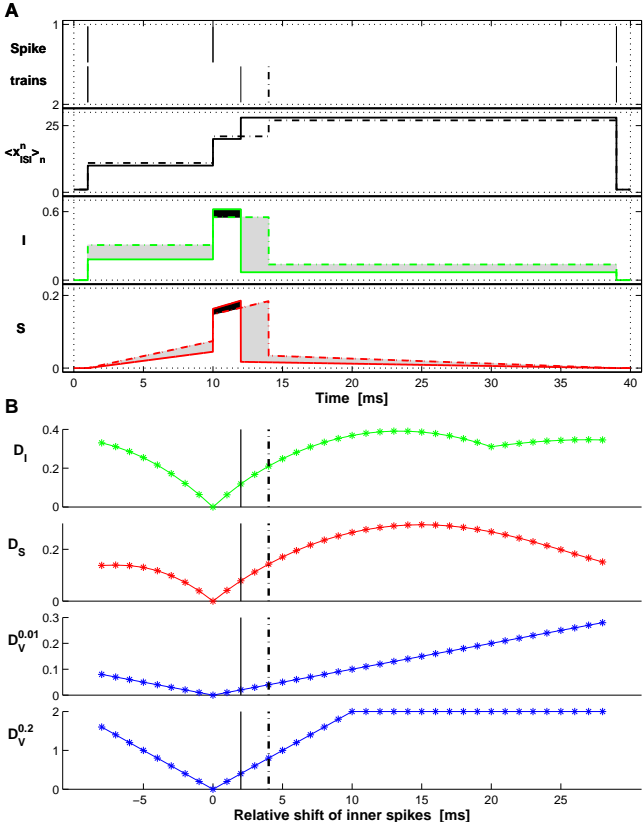}
    \caption{\abb\label{fig:Local-Behaviour} (color online) Local behaviour of SPIKE-distances. A. All spikes are set to fixed positions except for the inner spike of the second spike train which is shifted relative to the inner spike of the first spike train. In this example this spike is shifted from $12$ ms (solid lines) to $14$ ms (dashed lines). Following the spike trains and the mean instantaneous ISI-values we depict the instantaneous ISI-ratio $I (t)$ and the instantaneous spike time difference $S (t)$. The positive and negative deviations between these two profiles are marked by bright and dark shaded areas, respectively. In this example these two opposing contributions sum up to an increase of the distances.   B. Dependence of the ISI distance $D_{\mathrm {I}}$, the SPIKE-distance $D_{\mathrm {S}}$ and the Victor-Purpura distance $D_{\mathrm {V}}$ (for cost parameters $c_{\mathrm {V}}$ equal to $0.01$ and $0.2$) on the relative shift of the two inner spikes. The relative positions of the spikes in A are marked by vertical lines.}
\end{figure}

In order to illustrate the local behavior of the time-resolved ISI- and SPIKE-distances and compare it with the global behavior of the Victor-Purpura distance (cf. Appendix \ref{App-ss:Victor}), we start with a simple example (Fig. \ref{fig:Local-Behaviour}) composed of two identical spike trains with just three spikes each. Then, while all other spikes are kept fixed, the inner spike of the second spike train is shifted in time.

Exemplary spike trains and the results for the time-resolved bivariate measures $I (t)$ and $S (t)$ are shown in Fig. \ref{fig:Local-Behaviour}A. Both measures show local signatures that are somehow counter-intuitive: They assign larger positive deviations for more closer spikes. The reason is as follows: As the difference between the two inner spikes increases, the value of the instantaneous dissimilarities in the interval between these spikes (the peak height of the event kernel) decreases, however, at the same time the support interval of the peak value is getting longer as are the dissimilarity values in the remaining intervals (the drop height of the event kernel) although their support is getting smaller. Initially the sum effect is positive (compare the dark and the bright areas), however, as the difference between the two spikes is getting even larger, the relative importance of these two opposing effects changes. Accordingly, the corresponding distances $D_{\mathrm {I}}$ and $D_{\mathrm {S}}$ shown in Fig. \ref{fig:Local-Behaviour}B exhibit an initial increase which then leads to a maximum value for an intermediate shift followed again by a decrease. As the different slopes of the initial increase in distance for positive and negative shifts prove, the relative importance of the two opposing effects does not only depend on the shift between the two intermediate spikes, but is also influenced by the relative lengths of the outer intervals.

One peculiarity can be observed for the ISI-distance $D_{\mathrm {I}}$, a local minimum for a shift of $20$ ms. This is the point for which the two spike trains are symmetric under time reversal. Accordingly, in the central interval between the two inner spikes the instantaneous measure $I (t)$ equals zero and this leads to the low value of $D_{\mathrm {I}}$.

In Fig. \ref{fig:Local-Behaviour}B we also show results of the Victor-Purpura distance $D_{\mathrm {V}}$ for two cost values $c_{\mathrm {V}} = 0.01$ and $c_{\mathrm {V}} = 0.2$. A linear increase can be observed with the slope of the increase given by the cost parameter. Note that no distance values higher than $2$ can be obtained, since for high costs and large differences of the two inner spikes shifting gets too expensive and it becomes more affordable to delete and to insert the spike for a cost of $2$.

The difference between the instantaneous measures and the Victor-Purpura distance stems from the fact that the first are local and the latter is global. For the SPIKE-distance the inner spike of the second spike train is approaching the third spike of the first spike train which leads to the effect that the two spike trains are getting more similar
again. The Victor-Purpura distance on the other hand matches the two third spikes and thus considers only the difference between the two inner spikes although they are further apart than the later spikes.

\section{\label{s:Results} Results}

\subsection{\label{ss:Results-1-Capability} Simulation: Capability to distinguish different levels of spike train synchrony}

As a first quantitative test for the different measures we evaluated whether they are able to track continuous changes in synchrony when the transition to synchrony is based on coincidences of spike times. Measures comprise the bivariate and multivariate ISI- and SPIKE-distances $D_{\mathrm {I}}^{\mathrm {a}}$, $D_{\mathrm {I}}^{\mathrm {m}}$, $D_{\mathrm {S}}^{\mathrm {a}}$, and $D_{\mathrm {S}}^{\mathrm {m}}$, as well as the Victor-Purpura distance $D_{\mathrm {V}}$ and the correlation coefficient $C$ (cf. Appendix \ref{App-ss:Corrcoef}). The latter two measures are both represented by two parameter values, one of which was optimized over different values of the respective parameter.

We generated $25$ spike trains with $100$ spikes each. The spikes were randomly distributed within the interval $[0, 1]$ using a sampling interval of $0.0001$. In order to create spike trains with increasing levels of correlations we followed a correlation scheme introduced in \citet{Kreuz06} and defined a matching parameter $m$ that governed the fraction of shared spikes for each pair of neurons: for matching $m = 0$ all spikes are randomly distributed, whereas for matching $m = 1$ all spike trains are identical.

\begin{figure}
    \includegraphics[width=85mm]{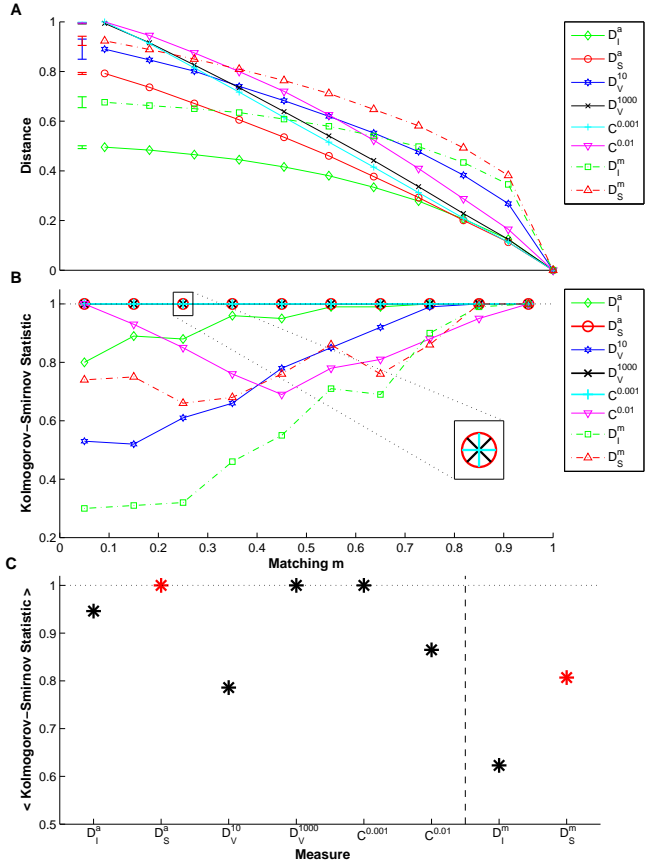}
    \caption{\abb\label{fig:Multi-KS-Comp} (color online) Capability to distinguish different levels of spike train synchrony.  A. ISI- and SPIKE-distances as well as Victor-Purpura distance (normalized to its maximum value) and correlation coefficient versus the matching parameter. Values shown are averages and standard deviations of distributions from $100$ realizations with $N = 25$ spike trains and $M = 100$ spikes each. For clarity we only show the standard deviations for matching $m = 0$, the other standard deviations decrease rather monotonously with the matching.   B. Kolmogorov-Smirnov statistics showing the distinguishability of distributions with neighboring matchings. Values for the different measures are plotted against the mean matching of the respective two distributions. C. Comparison of measures: Kolmogorov-Smirnov statistics averaged over all neighboring matchings. The vertical line separates the averaged bivariate and the multivariate measures.}
\end{figure}

Average values and standard deviations over $100$ realizations are shown in Fig. \ref{fig:Multi-KS-Comp}A. All distances equal $0$ for a perfect matching ($m = 1$). For decreasing matchings all average values increase monotonically, however, the shape of the increase differs. The multivariate methods first show a very steep and then a rather moderate increase. The averaged bivariate methods exhibit a more gradual increase except for the averaged bivariate ISI-distance whose increase is rather slow for low matchings. Furthermore, as indicated by the standard deviations of the distributions, the averaged bivariate methods are characterized by considerably narrower distributions. The average bivariate SPIKE-distance $D_{\mathrm {S}}^{\mathrm {a}}$, the optimized Victor-Purpura distance $D_{\mathrm {V}}$, and the optimized correlation coefficient $C$ are the only measures to combine a rather constant increase with a low variability.

Similar to the analysis of \citet{Kreuz09a} (see also Appendix \ref{App-ss:HR-Multivariate-Comparison}) we quantified the capability of the measures to distinguish distributions with neighboring matchings using Kolmogorov-Smirnov statistics (Fig. \ref{fig:Multi-KS-Comp}B). For the averaged bivariate SPIKE-distance as well as for the optimized optimized Victor-Purpura distance and the optimized correlation coefficient the maximum value of $1$ is obtained indicating pairs of non-overlapping neighboring distributions. For the other measures deviations from $1$ indicate overlapping distributions. This is quantified by the average of the Kolmogorov-Smirnov statistics over all matching transitions (Fig. \ref{fig:Multi-KS-Comp}C). Perfect scores are obtained for the averaged bivariate SPIKE-distance, the optimized Victor-Purpura distance, and the optimized correlation coefficient followed by the averaged bivariate ISI-distance. In general, the multivariate measures yield lower averages with the ISI-distance again lagging behind the SPIKE-distance.

\subsection{\label{ss:Results-2-Application-to-Real-Data} Application to real data}

In Section \ref{sss:Results-3-Application-to-RGCs} we investigate the spike train similarity within a complete population of retinal ganglion cells, while in Section \ref{sss:Results-3-Application-to-StimDis} we evaluate how well the various measures can discriminate single-unit responses to different taste stimuli. Since in both applications we use setups which involve only similarities between pairs of spike trains, we only compare the basic bivariate measures and not the averaged bivariate or multivariate extensions. Thus the measures are narrowed down to the ISI-distance $D_{\mathrm {I}}$, the SPIKE-distance $D_{\mathrm {S}}$, the Victor-Purpura distance $D_{\mathrm {V}}$~\footnote{Note that among the time-scale dependent spike distances we restrict ourselves to the Victor-Purpura distance, since this was the measure that proved to be the best performer in previous measure comparisons (\citet{Kreuz07c, Kreuz09a}, see also Appendix \ref{App-s:HR-Comparison}).}, and the correlation coefficient $C$. In both applications for $D_{\mathrm {V}}$ and $C$ we sampled the relevant range of the respective time-scale parameter equidistantly on a logarithmic scale.

\subsubsection{\label{sss:Results-3-Application-to-RGCs} Spike train similarity among retinal ganglion cells}

\begin{figure}
    \includegraphics[width=85mm]{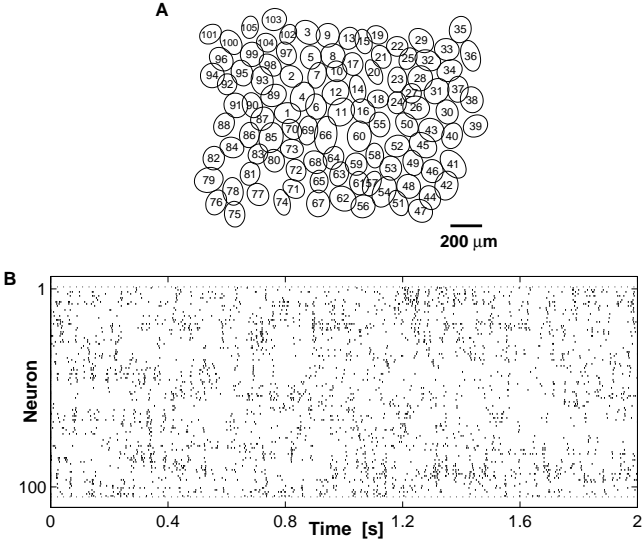}
    \caption{\abb\label{fig:Martin-Data} A. Receptive fields (Gaussian fit outlines) of an almost complete population of $N = 105$ ON-Parasol retinal ganglion cells. This is a reanalyzed subset of the first dataset of \citet{Shlens06, Shlens09}.  B. Spike trains from these $105$ cells during the first two seconds of one realization of white noise stimulation.}
\end{figure}

We first applied the bivariate spike train distances to responses from an almost complete population ($N = 105$ neurons) of ON parasol retinal ganglion cells (\citeauthor{Shlens06}, \citeyear{Shlens06}, \citeyear{Shlens09}; for more details on the data confer Appendix \ref{App-ss:Data-Retinal-Ganglion-Cells}). These recordings were performed using both white noise stimulation and constant, spatially uniform illumination. The white noise stimulation was also used to identify the receptive field of the retinal ganglion cells by means of reverse correlation \cite{Chichilnisky01}. As can be seen in Fig. \ref{fig:Martin-Data}A, the receptive fields of the complete population form an orderly mosaic that tiles visual space with minimal overlap. An exemplary snapshot of the spiking activity of the population responding to a white noise stimulation is shown in Fig. \ref{fig:Martin-Data}B.

From this mosaic we extracted the pairwise distances between the centers of the respective fields. The amount of synchronized firing between pairs of RGCs of the same type declines systematically with distance between the two cells \citep{Mastronarde83, Meister95, DeVries99, Shlens06, Greschner10}. This well described behavior was used as a benchmark against which we tested the capability of the various spike train distances to assign higher values to more distant cell pairs (or lower values to more distant cell pairs in case of the correlation coefficient $C$). Note that the distance between cell pairs does not explain all the variance in the amount of synchronized firing and that this behavior was mainly studied using measures related to the correlation coefficient.
\begin{figure}
    \includegraphics[width=85mm]{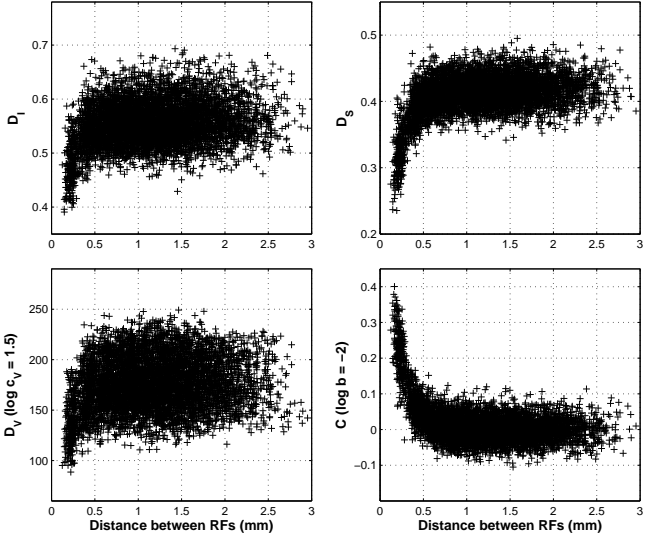}
    \caption{\abb\label{fig:Martin-Master-Scatter-Plots} Scatter plots of the ISI-distance $D_{\mathrm {I}}$, the SPIKE-distance $D_{\mathrm {S}}$, the Victor-Purpura distance $D_{\mathrm {V}}$ (for $\log c_{\mathrm {V}} = 1.5$), and the correlation coefficient $C$ (for $\log b = -2$) versus the distance between the centers of the receptive fields (RFs). Note that $C$ is a measure of similarity and thus exhibits a decrease with the distance between the RFs. For both $D_{\mathrm {V}}$ and $C$ we show the plots for the parameter value that yielded the highest absolute Spearman's rank correlation coefficient $R$ (cf. Fig. \ref{fig:Martin-Master-Grey-Comp-Spearman}). The values of $R$ for this example were $0.23$ ($D_{\mathrm {I}}$), $0.36$ ($D_{\mathrm {S}}$), $0.17$ ($D_{\mathrm {V}}$), and $0.39$ ($C$), respectively.}
\end{figure}

For each of these measures and for both kinds of stimulations we first calculated all pairwise spike train distances for the complete population. Exemplary scatter plots of the measures versus the distance between the receptive fields are shown for one segment of the white noise stimulation in Fig. \ref{fig:Martin-Master-Scatter-Plots}. All measures exhibit a pronounced saturation for higher distances between RFs but show a very clear dependency for lower distances. However, the measures differ in how pronounced this dependency is relative to the overall variance. While it is rather low for the Victor-Purpura distance and the ISI-distance, it is much larger for the SPIKE-distance and the correlation coefficient.

\begin{figure}
    \includegraphics[width=85mm]{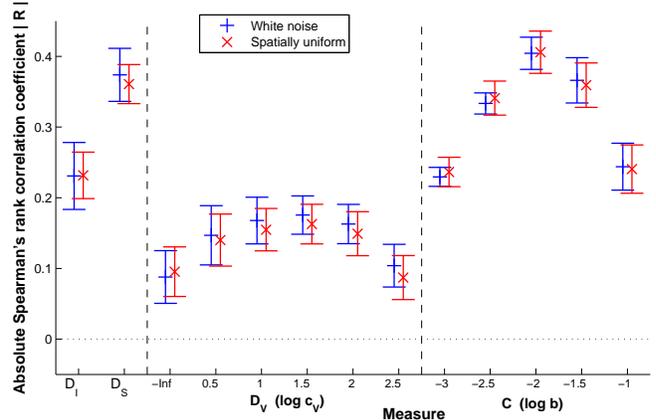}
    \caption{\abb\label{fig:Martin-Master-Grey-Comp-Spearman} Comparison of the absolute Spearman's rank correlation coefficients (average and standard deviation) over $18$ segments ($10$ s each) for both white noise stimulation and constant, spatially uniform illumination. Measures comprise the ISI- and the SPIKE-distance, the Victor-Purpura distance, and the correlation coefficient. The parameters of the latter two measures are varied on a logarithmic scale.}
\end{figure}

For all measures we quantified the relation between the spike train distances and the distances between the receptive fields using the absolute Spearman's rank correlation coefficient $R$ (Fig. \ref{fig:Martin-Master-Grey-Comp-Spearman}). For both the white noise stimulation and the constant, spatially uniform illumination the highest average $R$-values are obtained for the optimized correlation coefficient $C$. This is closely followed by the SPIKE-distance $D_{\mathrm {S}}$ which in turn yields a considerable improvement over the ISI-distance $D_{\mathrm {I}}$. For the Victor-Purpura distances $D_{\mathrm {V}}$ despite the optimization the highest coefficient is considerably lower than the values obtained for the other methods. Also results are not very robust since a more detailed investigation revealed that for different segments the actual maximum $R$-value was attained for cost values covering two entire logarithmic decades.

\subsubsection{\label{sss:Results-3-Application-to-StimDis} Taste stimuli discrimination from single-unit recordings}

As a second application to real data we analyze one complete dataset from "neurodatabase.org" \citep{Gardner04}. It consists of single-unit responses of three neurons ($\#4$, $\#9$, and $\#11$) in the nucleus of the solitary tract (NTS) of rats to four different taste stimuli \citep{DiLorenzo03}. For the three different neurons we have $N_R = 19$, $23$, and $16$ repetitions, respectively, for each stimulus. For more details on the data confer Appendix \ref{App-ss:Data-Victor}.

For all measures and for each neuron separately we calculate the pairwise distance matrix for all $4 \cdot N_R$ realizations. Assuming stimulus-driven synchronization, a good discrimination should yield low distance values for responses to the same stimuli and high distance values to responses for different stimuli (and vice versa for the correlation coefficient).

In order to evaluate the separation of the spike trains into $4$ different clusters, we follow \citeauthor{Victor96} and  define the distance between the spike train $S_i$ and the cluster $C_\alpha$ as $\langle d(S_i,S_j) \rangle_\alpha$, where $\langle \cdot \rangle_\alpha$ denotes the average over all spike trains in the cluster $C_\alpha$. From these distances we compute the normalized confusion matrix $p_{\alpha \beta}$ (\citeauthor{Abramson63}, \citeyear{Abramson63}; \citeauthor{Victor96}, \citeyear{Victor96}) whose entry $p_{\alpha\beta}$ is defined as the probability that $C_\beta$ is the closest cluster to a spike train belonging to $C_\alpha$. For a perfect clustering the confusion matrix is diagonal, whereas each misclassification increments a non-diagonal element. The performance of the measures in discriminating the different stimuli is finally quantified by the normalized mutual information
\begin{equation}\label{eq:Entropy}
I_C = \left ( \sum_{\alpha,\beta} p_{\alpha\beta} \log \frac{p_{\alpha\beta}}{P_\alpha P_\beta} \right ) /I_{max}
\end{equation}
where $P_\alpha = \sum_b p_{\alpha b}$, and $P_\beta = \sum_b p_{b\beta}$ and $I_{max}$ denotes the maximum mutual information obtained for a correct classification.

\begin{figure}
    \includegraphics[width=85mm]{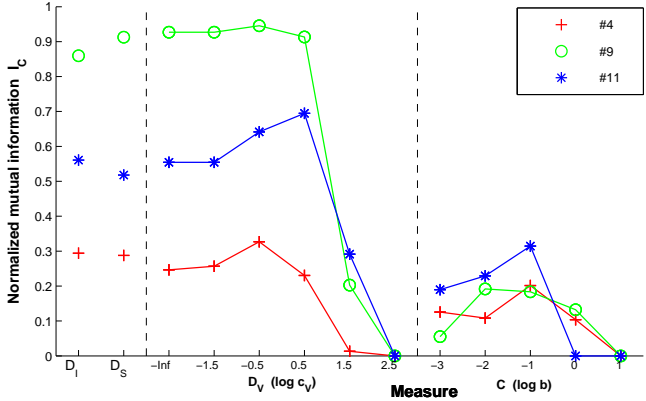}
    \caption{\abb\label{fig:Victor-Comp-DiLorenzo} Normalized confusion mutual information for three neurons responding to different taste stimuli. The same measures as in Fig. \ref{fig:Martin-Master-Grey-Comp-Spearman} were evaluated, only the range of the time-scale parameters was adapted in order to cover the maxima of discrimination for this task.}
\end{figure}

Results for the three neurons recorded during the taste stimuli discrimination task \citep{DiLorenzo03} are shown in Fig. \ref{fig:Victor-Comp-DiLorenzo}. For all three neurons the highest discrimination is obtained for the Victor-Purpura distance, however, as already reported in \citet{DiLorenzo03} for different neurons the best discrimination is obtained for different coding schemes. The neurons $\#4$ and $\#11$ exhibit their maximum for positive values of the cost parameter indicating that a temporal coding scheme allow for greater discrimination of the four stimuli. Only for neuron $\#9$ the curve does not increase considerably for positive cost parameters. In this case the spike timing is less relevant and a rate coding scheme is sufficient for a proper discrimination of the tastants. Slightly below the maximum $I_C$-values obtained for the Victor-Purpura distance follow the time-scale adaptive ISI-distance and SPIKE-distance. In this task there seems to be no clear distinction between their discrimination values regardless of the neuron's coding scheme. Finally, for all three neurons the lowest discrimination capabilities are obtained by the correlation coefficient.

\section{\label{s:Discussion} Discussion}

Motivated by the prevalence of coincident spiking in many different areas of the brain, we propose a SPIKE-distance which, together with its multi-neuron extensions, can be applied to quantify the (dis)similarity of two or more spike trains. These measures are sensitive to the timing of spikes and are thus complementary to the ISI-distance \citep{Kreuz07c} and its extensions \citep{Kreuz09a} which are instead sensitive to the relative lengths of simultaneous interspike intervals.

At the same time the SPIKE-distance and its extensions share many distinct properties with the ISI-distance, in particular, they are parameter free and time-scale adaptive. The latter property can be preferable in applications to real data for which there is no validated knowledge about the relevant time scales. While time-scale dependent measures yield a functional characterization of the spike trains, a single valued method gives a more objective and comparable estimate of neuronal variability. Other drawbacks of time-scale dependent measures are the computational cost and the time and effort that is needed to find the right parameter. Moreover, it is not at all guaranteed that there exists a right parameter. For example, spike trains that include different time-scales such as regular spiking and bursting might result in misleading conclusions, since any fixed parameter will misrepresent either one of those dynamics.

One of the main arguments for the use of a time-scale dependent measures is their potential insight into the precision of the neuronal code \citep{Victor96}. A very common setup in experimental neuroscience is the recording of single- or multi-unit spike train responses for the repeated presentation of different classes of stimuli (for an overview see \citeauthor{Victor05}, \citeyear{Victor05}). The pairwise distance matrices over all realizations are calculated and used as input for a cluster analysis of the type used in Section \ref{sss:Results-3-Application-to-StimDis}. The time-scale for which the responses to different stimuli are best discriminated is then assumed to be the precision of the neural code. However, a recent investigation of this kind of precision analysis \citep{Chicharro11} has shown that the optimal time-scale obtained from the cluster analysis is far from being conclusive, but rather results in a non-trivial way from the interplay of different factors. These factors include the distribution of the information contained in different parts of the response and the degree of redundancy between them. For dynamic stimuli the optimal time-scale also depends on the stimuli statistics, in particular, the temporal distribution of the stimulus features to which the neurons respond.

Despite the problems in the interpretation of the optimal time-scale, the Victor-Purpura distance is designed to test for specific codes ranging from a rate code to a coincidence detector \citep{Victor97}. The correlation coefficient focusses purely on the coincidences and thus reflects the most specific coding assumption. On the other extreme, the most general approach is taken by the ISI-distance and the SPIKE-distance which, in contrast to the Victor-Purpura distance, are time-scale adaptive and parameter-free. Thus the different measures can be ordered from measures tied to a specific hypothesis about the code (correlation coefficient) via intermediate measures (Victor-Purpura distance) to measures that are very general (ISI- and SPIKE distance). Measures on different ends of this scale are complementary in nature. The ones capable of quantifying the degree of dissimilarity of the spike trains without relying on specific coding hypotheses are very well suited for an exploratory analysis. Oppositely, once a particular coding scheme is hypothesized more specific measures are needed for a confirmatory analysis.

In the application to the retina data presented in Section \ref{sss:Results-3-Application-to-RGCs} the correlation coefficient, which is focussed on spike coincidences, yields best results, i.e., the highest absolute Spearman's rank correlation coefficients. On the other hand, in Section \ref{sss:Results-3-Application-to-StimDis} the same measure fails to discriminate the different taste stimuli. The former result is consistent with the prevalence of synchronized firing in the retina \citep{Shlens08}, while in the latter taste discrimination task the distinction seems to rely on coding schemes other than coincidences \citep{DiLorenzo03}. The more general ISI- and SPIKE-distances show a very adaptive behavior and yield a good performance in both applications. It should be noted that while both measures discriminate the taste stimuli about equally well, the SPIKE-distance outperforms the ISI-distance on the retina data. This is consistent with the previous observation: When synchronized firing is prevalent, the sensitivity to spike coincidences exhibited by the SPIKE-distance becomes relevant and leads to a considerable improvement with respect to the ISI-distance.

Both ISI- and SPIKE-distance are conceptually simple, computationally efficient and easy to visualize in a time-resolved manner. This kind of visualization can be achieved because the distances are defined as temporal averages over time profiles, i.e., they are based on instantaneous values which are calculated from the sequences of spike times. However, it should be noted that each instantaneous value cannot be interpreted without looking at its local context. The SPIKE-distances are time-resolved only in the sense that they assign to each spiking event an event kernel. In case the focus of attention is on the long-term behavior, these local event kernels can easily be eliminated by appropriate moving averaging (as shown in Fig. \ref{fig:Constructed-Example}b). Another caveat regarding all measures of spike train synchrony is that they generally assume a zero time delay between spike trains. Accordingly, in case of a non-zero time delay, e.g., caused by finite signal transmission times between different brain regions, it should be detected and eliminated beforehand \citep{Waddell07, Nawrot03}.

The measures used in this study quantify the level of synchrony within one set of two or more spike trains. For the Victor-Purpura and the van Rossum distances (\citeauthor{VanRossum01}, \citeyear{VanRossum01}, cf. Appendix \ref{App-s:HR-Comparison}) there also exist extensions that estimate the synchrony between two populations of neurons (\citeauthor{Aronov03}, \citeyear{Aronov03} and \citeauthor{Houghton08}, \citeyear{Houghton08}, respectively). Corresponding extensions for both the ISI- and the SPIKE-distance will be presented in a forthcoming study \citep{Kreuz11}.

We close by pointing out once more that the Matlab source code for calculating and visualizing both the ISI- and the SPIKE-distances can be downloaded at www.fi.isc.cnr.it/users/thomas.kreuz/sourcecode.html.



\begin{appendix} \label{s:Appendix}

\section{\label{App-s:SPIKE-distance} Derivation of the SPIKE-distance}

The derivation of the instantaneous SPIKE-measure $S (t)$ (Eq. \ref{eq:Bi-Spike-Diss2}) consists of three steps: calculation of instantaneous spike time differences, locally weighted averaging, and normalization. Here we provide additional details about the motivation behind each of these steps.

The first step is the calculation of the absolute instantaneous differences between previous and following spike times $| \Delta t_{\mathrm {P}} (t) |$ and $| \Delta t_{\mathrm {F}} (t) |$ (Eqs. \ref{eq:Prev-Diff} and \ref{eq:Foll-Diff}, respectively). By taking into account only the previous and the following spike in each spike train the method relies on local information only. The method is also time-scale adaptive, since the information used is not contained within a window of fixed size but rather within a time frame whose size depends on the local rate of each spike train.

The next two steps, locally weighted average, and normalization are interchangeable. Here we first calculate the locally weighted average of the two differences $| \Delta t_{\mathrm {j}} (t) |, j = \mathrm {P},\mathrm {F}$:
\begin{equation} \label{eq:App-Weighted-Mean}
    \langle \Delta t_{\mathrm {j}} (t) \rangle_{j=\mathrm {P},\mathrm {F}} = \frac{\sum_{j=\mathrm {P},\mathrm {F}} | \Delta t_{\mathrm {j}} (t)| f(x_{\mathrm {j}}^{(n)} (t))}{\sum_{j=\mathrm {P},\mathrm {F}} f(x_{\mathrm {j}}^{(n)} (t))}.
\end{equation}
Using the inverse of the mean intervals from the time instant under consideration to the previous and the following spikes $\langle x_{\mathrm {j}}^{(n)} (t) \rangle, j = \mathrm {P},\mathrm {F}$ (averages over Eqs. \ref{eq:Prev-Spike-Dist} and \ref{eq:Foll-Spike-Dist}, respectively) as weights we obtain
\begin{equation} \label{eq:App-Weighted-Mean-Detailed}
    \langle \Delta t_j (t) \rangle_{j=\mathrm {P},\mathrm {F}} = \frac{| \Delta t_{\mathrm {P}} (t) | \frac{1}{\langle x_{\mathrm {P}}^{(n)} (t)\rangle_n}+| \Delta t_{\mathrm {F}} (t) | \frac{1}{\langle x_{\mathrm {F}}^{(n)} (t) \rangle_n}}{\frac{1}{\langle x_{\mathrm {P}}^{(n)} (t)\rangle_n}+\frac{1}{\langle x_{\mathrm {F}}^{(n)} (t)\rangle_n}}
\end{equation}
which can easily be transformed into
\begin{equation} \label{eq:App-Weighted-Mean-Basic}
    \langle \Delta t_j (t) \rangle_{j=\mathrm {P},\mathrm {F}} = \frac{| \Delta t_{\mathrm {P}} (t) | \langle x_{\mathrm {F}}^{(n)} (t)\rangle_n + | \Delta t_{\mathrm {F}} (t) | \langle x_{\mathrm {P}}^{(n)} (t)\rangle_n }{\langle x_{\mathrm {P}}^{(n)} (t)\rangle_n + \langle x_{\mathrm {F}}^{(n)} (t)\rangle_n}.
\end{equation}
As can be seen in Fig. \ref{fig:Spike-Distance-Example}, the sum of the mean intervals equals the mean interspike interval:
\begin{equation} \label{eq:App-Mean-Prev-Foll-Spike}
    \langle x_{\mathrm {P}}^{(n)} (t) \rangle_n + \langle x_{\mathrm {F}}^{(n)} (t) \rangle_n = \langle x_{\mathrm {ISI}}^{(n)} (t) \rangle_n.
\end{equation}
Inserting Eq. \ref{eq:App-Mean-Prev-Foll-Spike} into Eq. \ref{eq:App-Weighted-Mean-Basic} we obtain
\begin{equation}
    \langle \Delta t_j (t) \rangle_{j=\mathrm {P},\mathrm {F}} = \frac{| \Delta t_{\mathrm {P}} (t) | \langle x_{\mathrm {F}}^{(n)} (t)\rangle_n + | \Delta t_{\mathrm {F}} (t) | \langle x_{\mathrm {P}}^{(n)} (t)\rangle_n }{\langle x_{\mathrm {ISI}}^{(n)} (t) \rangle_n}.
\end{equation}

Finally, in order to achieve time-scale invariance (i.e., all stretched and compressed spike trains should yield the same value) we divide this locally weighted spike time difference by the mean interspike interval. This way we recover Eq. \ref{eq:Bi-Spike-Diss2}:
\begin{equation}
    S (t) = \frac{| \Delta t_{\mathrm {P}} (t) | \langle x_{\mathrm {F}}^{(n)} (t)\rangle_n + | \Delta t_{\mathrm {F}} (t) | \langle x_{\mathrm {P}}^{(n)} (t)\rangle_n }{\langle x_{\mathrm {ISI}}^{(n)} (t) \rangle_n^2}.
\end{equation}
At the same time the spike time differences are related to the local spike rate such that a certain spike time difference is the more relevant the higher the rate. This also yields to a normalization. The maximum value $1$ of $S (t)$ is approached when two spikes from the different spike trains almost coincide. This can be seen when looking at Fig. 1B and imagining $t_{\mathrm {P}}^{(1)}$ getting closer and closer to $t_{\mathrm {F}}^{(2)}$. In this case $| \Delta t_{\mathrm {P}} (t) |$ approaches $x_{\mathrm {ISI}}^{(2)}$ and $| \Delta t_{\mathrm {F}} (t) |$ approaches $x_{\mathrm {ISI}}^{(1)}$. Accordingly, $S (t)$ approaches the value $1$. Note that overall the local increase is outbalanced by surrounding decreases so that for converging spikes a reduction of the SPIKE-distance is obtained (cf. Section \ref{ss:Methods-3-Local-behaviour}). Also note that, while $S (t)$ can get arbitrarily close to $1$ it never actually reaches this value, because for identical spikes $0$ is obtained.
Apart from the normalization, an analogous derivation holds for the multivariate case (Section \ref{sss:Multivariate-Spike-Distance}) as well. For both cases the distances are obtained from the instantaneous values as their temporal average (Eqs. \ref{eq:Bi-Spike-Distances} and \ref{eq:Multi-Spike-Distances}).


\section{\label{App-s:Existing-Methods} Previously published measures of spike train distance}

Here we restrict ourselves to descriptions of the ISI-distance together with its extensions, the Victor-Purpura distance, and the correlation coefficient since these are the measures that we deal with in more detail. For the other measures that we compare against in Appendix \ref{App-s:HR-Comparison} please refer to the Appendix of \citet{Kreuz09a}.

\subsection{\label{App-ss:ISI-Distances} The bivariate ISI-distance and its extensions}

The ISI-distance \citep{Kreuz07c} and its extensions \citep{Kreuz09a} are based on the instantaneous interspike intervals.

\subsubsection{\label{App-sss:Bivariate-ISI-Distance} Bivariate ISI-distance}

To define a time-resolved, symmetric, and time-scale adaptive measure of the relative firing rate pattern \citep{Kreuz07c} we take the instantaneous ratio between $x_{\mathrm {ISI}}^{(1)}$ and $x_{\mathrm {ISI}}^{(2)}$ (Eq. \ref{eq:ISI}), and combine according to:
\begin{equation} \label{eq:ISI-Ratio}
    I_{12} (t) = \begin{cases}
           x_{\mathrm {ISI}}^{(1)} (t) / x_{\mathrm {ISI}}^{(2)} (t) - 1 & {\rm if} ~~ x_{\mathrm {ISI}}^{(1)} (t) \leq x_{\mathrm {ISI}}^{(2)} (t) \cr
                      - (x_{\mathrm {ISI}}^{(2)} (t) / x_{\mathrm {ISI}}^{(1)} (t) -1)     & {\rm otherwise}.
                  \end{cases}
\end{equation}
This quantity becomes $0$ for identical ISI in the two spike trains, and approaches $-1$ and $1$, respectively, if the first or the second spike train is much faster than the other.

The bivariate ISI-distance is obtained by averaging the absolute ISI-ratio over time:
\begin{equation} \label{eq:ISI-distance}
    D_{\mathrm {I}} = \frac{1}{T} \int_{t=0}^T dt | I_{12} (t) |.
\end{equation}
The bivariate ISI-distance is bounded in the interval $[0, 1]$.

\subsubsection{\label{App-sss:Averaged-ISI-Distance} Averaged bivariate ISI-distance}

As in the case of the averaged bivariate SPIKE-distance (Section \ref{sss:Averaged-Spike-Distance}) the averages over pairwise distances and over time commute and we can
visualize the instantaneous average in a time-resolved way. First we calculate the instantaneous average $A (t)$ over all pairwise absolute ISI-ratios $|I_{mn} (t)|$ (cf. Eq.
\ref{eq:ISI-Ratio})
\begin{equation} \label{eq:Instantaneous-ISI-Average}
    I^{\mathrm {a}} (t) = \frac{1}{N(N-1)/2}\sum_{m=1}^N \sum_{n=m+1}^N | I_{mn}(t) |
\end{equation}
before we average over time:
\begin{equation} \label{eq:Multi-ISI-distance}
    D_{\mathrm {I}}^{\mathrm {a}} = \frac{1}{T} \int_{t=0}^T dt I^{\mathrm {a}} (t).
\end{equation}
As the original measure, the averaged bivariate ISI-distance is restricted to the interval $[0, 1]$.

\subsubsection{\label{App-sss:Multivariate-ISI-Distance} Multivariate ISI-distance}

We derive a multivariate measure by calculating the instantaneous coefficient of variation taken across all neurons at any given instant in time
\begin{equation} \label{eq:ISI-CV}
    I^{\mathrm {m}} (t) = \frac{\sigma[x_{\mathrm {ISI}}^{(n)} (t)]_n }{\langle x_{\mathrm {ISI}}^{(n)} (t) \rangle_n}
\end{equation}
and averaging over time:
\begin{equation} \label{eq:Multi-Intra-ISI-distance}
    D_{\mathrm {I}}^{\mathrm {m}} = \frac{1}{T} \int_{t=0}^T dt I^{\mathrm {m}} (t).
\end{equation}
For identical spike trains $D_{\mathrm {I}}^{\mathrm {m}}$ obtains the same lower bound of zero as $D_{\mathrm {I}}$ and $D_{\mathrm {I}}^{\mathrm {a}}$, but unlike them it lacks an upper bound.

\subsection{\label{App-ss:Victor} Victor-Purpura distance}

The spike train distance $D_{\mathrm {V}}$ introduced in \citet{Victor96} defines the distance between two spike trains in terms of the minimum cost of transforming one spike train into the other by using just three basic operations: spike insertion, spike deletion and spike shift. While the cost of insertion or deletion of a spike is set to one, the cost $c_{\mathrm {V}}$ of moving a spike by some interval is the only parameter of the method, and sets the time-scale of the analysis. For zero cost, the distance is equal to the difference in spike counts, for high costs, the distance approaches the number of non-coincident spikes, as it becomes more favorable to delete all non-coincident spikes than to shift them. Thus, by increasing the cost, the distance is transformed from a rate distance to a timing distance. For multi-neuron data, synchrony is defined as the average over all pairs of spike trains:
\begin{equation} \label{eq:Averaged-Victor}
    D_{\mathrm {V}}^{\mathrm {a}} = \frac{1}{N(N-1)/2}\sum_{m=1}^N \sum_{n=m+1}^N D_{\mathrm {V}}^{mn}.
\end{equation}

\subsection{\label{App-ss:Corrcoef} Correlation coefficient}

In contrast to all the other measures used in this paper the bivariate correlation coefficient relies on binning with the bin size $b$ (in seconds) as a free parameter. If there is no spike within the bin, a zero is assigned, otherwise this bins' value is set to one~\footnote{Note that this binary setup of the correlation coefficient disregards all differences within the bin size as well as all differences larger than the bin size (similar to a rectangular kernel), whereas both the Victor-Purpura and the SPIKE-distance weight all spike time differences linearly (similar to a triangular kernel).}. The two resulting vectors $\overrightarrow{b}_x$ and $\overrightarrow{b}_y$ are used as input for the correlation coefficient $C$ which equals the normalized covariance function $c$ according to
\begin{equation} \label{eq:Correlation-coefficient}
    C = c(\overrightarrow{b}_x,\overrightarrow{b}_y)/\sqrt{c(\overrightarrow{b}_x,\overrightarrow{b}_x) \cdot c(\overrightarrow{b}_y,\overrightarrow{b}_y)}.
\end{equation}
Also the multi-neuron correlation coefficient is calculated by averaging over all pairs of spike trains.

\section{\label{App-s:Data} Data}

All recordings and simulations were performed prior to and independently from the design of this study.

\subsection{\label{App-ss:Data-Retinal-Ganglion-Cells} Multi-neuron recordings from retinal ganglion cells}

In Section \ref{sss:Results-3-Application-to-RGCs} the recorded activity of a nearly complete population of ON parasol retinal ganglion cells (RGCs) is used to compare the performance of the new SPIKE-distances with previously published measures such as the Victor-Purpura distance and the ISI-distances. The dataset used is a reanalyzed subset of the first dataset in \citet{Shlens06, Shlens09} (cf. Figs. 2A and 1A, respectively). RGCs were recorded with planar array of $512$ extracellular microelectrodes, covering an area of $1890 \times 900$ $\mu m$. The spikes of different cells were isolate \citep{Field07b} and ON parasol cells were identified by their density, their response kinetics, and their receptive field characteristics \citep{Chichilnisky02}. Responses to two kinds of stimulations were analyzed. Spatiotemporal receptive fields were measured using reverse correlation with a white noise stimulus, composed of a lattice of square pixels updating randomly and independently over time \citep{Chichilnisky01}. The intensity of each display primary at each pixel location was chosen independently from a binary distribution at each refresh. A stimulus pixel size of $60$ $\mu m$ on a side was used. Measurements of spontaneous activity were obtained in the presence of a spatially uniform, full-field illumination with intensity equal to the mean intensity of the white noise stimulus. For both stimuli, white noise stimulation and constant illumination, we analyzed $18$ segments of $10$ s each which were distributed regularly over a time scan of $30$ minutes.

\subsection{\label{App-ss:Data-Victor} Taste response and temporal coding in the nucleus of the solitary tract of the rat}

In Section \ref{sss:Results-3-Application-to-StimDis} we compare the performance of the spike train distances in discriminating four different taste stimuli. In \citet{DiLorenzo03} the authors examined the reliability of response rate across stimulus repetitions and the potential contribution of temporal coding to the discrimination of four taste stimuli (NaCl, sucrose, quinine HCl, and HCl) in the nucleus of the solitary tract (NTS) of rats. A selected dataset consisting of responses recorded from three different neurons is available online (http://neurodatabase.org/, \citeauthor{Gardner04}, \citeyear{Gardner04}) as experiment "dilorenzo-ndb-1181". For each of the four stimuli recordings of these cells, labeled $\#4$, $\#9$, and $\#11$, include $19$, $23$, and $16$ repetitions, respectively.

\subsection{\label{App-ss:Data-Hindemarsh-Rose} Hindemarsh-Rose simulations}

The spike trains of the controlled configuration used in Appendix \ref{App-s:HR-Comparison} were generated using time series extracted from a network of Hindemarsh-Rose neurons \citep{Hindmarsh84}. This network was originally designed to analyze semantic memory representations using feature-based models. Details of the network architecture and
the implementation of the feature coding can be found in \citet{Morelli05}. The clustering properties of these data were detailed in \citet{Kreuz07c}, their set separation was investigated in \citet{Kreuz09a}, and we here follow the description of these data given in \citet{Kreuz09a}.

In short, the network consisted of $128$ Hindemarsh-Rose neurons belonging to $U = 16$ different modules with $F = 8$ neurons each. The state of the neuron $n$ was determined by three dimensionless first-order differential equations describing the evolution of the membrane potential $X_n$, the recovery variable $Y_n$, and a slow adaptation current $Z_n$,
\begin{eqnarray} \label{eq:Hindemarsh-Rose}
    \dot{X_n} &=& Y_{n} - X_n^3 + 3X_n^{2} - Z_n + I_n + \alpha_n - \beta_n \\
    \dot{Y_n} &=& 1 - 5X_n^2 - Y_n \\
    \dot{Z_n} &=& 0.006 [4(X_n-1.6)-Z_n],
\end{eqnarray}
where
\begin{equation} \label{eq:HR-impulse-current}
    \alpha_n = \sum_{m=1}^{F(U-1)} w_{nm} A_m
\end{equation}
and
\begin{equation} \label{eq:HR-local-inhibition}
    \beta_n = \frac{1}{F-1} \sum_{m=1}^{F-1} A_m^{(n)}
\end{equation}
are the weighted inter-modular and intra-modular activities, respectively, which are dependent on the synaptic connection weights $w_{nm}$ and the neuronal activity variables $A_n$. The external input $I_n$ was chosen randomly between $3.0$ and $3.1$ such that the Hindemarsh-Rose neurons were operating in a chaotic regime.

In a learning stage, input memory patterns were stored by updating the synaptic connection weights $w_{nm}$ between different neurons using a Hebbian mechanism based on the activity variables $A_n$. A neuron was considered active whenever its membrane potential $X$ exceeded the threshold value $\widehat{X} = 0$ and its activity was coded by the variable $A_n = \theta(X_n - \widehat{X})$, where $\theta$ is the Heavyside function with $\theta(x) = 1$ if $x \geq 0$ and $\theta(x) = 0$ if $x < 0$. For these studies we restricted ourselves to $29$ time series $X_n$ extracted during the retrieval stage in which the learned connection weights were kept constant. According to their coding properties regarding the retrieval of only two distinguished memory patterns, they belonged to three clusters: $13$ of the neurons coded for pattern $1$ only, $13$ coded for pattern $2$ only, and $3$ coded for both pattern (``Shared''). The respective time series were labelled by ``1'', ``2'' and ``S'' followed by an index letter. The numerical integration was done using a fourth-order Runge-Kutta integration with a fixed step-size of $0.05$ (arbitrary time units). The length of the time series was $32768$ data points. The threshold for spike detection was chosen as the arithmetic average over the minimum and maximum value of the respective time series. In Appendix \ref{App-ss:HR-Bivariate-Comparison} we follow \citet{Kreuz07c} and evaluate the clustering of all $29$ spike trains, whereas in Appendix \ref{App-ss:HR-Multivariate-Comparison} we follow \citet{Kreuz09a} and the investigation of the set separations was restricted to the two principal clusters.

\section{\label{App-s:HR-Comparison} Comparison of measures using simulated Hindemarsh-Rose time series}

In order to connect with our previous work we finally evaluate the performance of the newly proposed SPIKE-distances within the bivariate and multivariate setups of the simulated network of Hindemarsh-Rose neurons used in \citet{Kreuz07c} and \citet{Kreuz09a}, respectively.

\subsection{\label{App-ss:HR-Bivariate-Comparison} Bivariate setup: Assessing clustering quality}

One important application for measures of spike train synchrony is the identification of spike train correlations and the recognition of response clusters. In \citet{Kreuz07c} we carried out a controlled comparison of six different spike train distances regarding their capability to reproduce the clustering within a network of $29$ Hindemarsh-Rose spike trains (for a description of the data see Appendix \ref{App-ss:Data-Hindemarsh-Rose}).

The measures against which we compared the ISI-distance $D_{\mathrm {I}}$ were the Victor-Purpura distance $D_{\mathrm {V}}$ \citep{Victor96}, the van Rossum metric $D_{\mathrm {R}}$ \citep{VanRossum01}, the inversion of the similarity measure proposed by \citet{S_Schreiber03}, and the inverted event synchronization $D_{\mathrm {Q}}$ \citep{QuianQuiroga02b}. The van Rossum metric $D_{\mathrm {R}}$ measures the Euclidean distance between the two spike trains after filtering of the spikes with an exponential kernel. The distance $D_{\mathrm {Sch}}$ based on the similarity measure proposed by \citet{S_Schreiber03} quantifies the normalized cross correlation of spike trains after Gaussian filtering. The inverted event synchronization $D_{\mathrm {Q}}$ \citep{QuianQuiroga02b} quantifies the number of quasi-simultaneous appearances using a variable time-scale that automatically adapts itself to the local spike rates. Thus $D_{\mathrm {Q}}$ \citep{QuianQuiroga02b}, like the ISI- and the SPIKE-distance, does not need a time-scale parameter, whereas all other distances rely on a parameter that sets the time-scale of the analysis. These parameters were varied over several orders of magnitude and the highest performance obtained was used to represent the measure.

\begin{figure}
    \includegraphics[width=85mm]{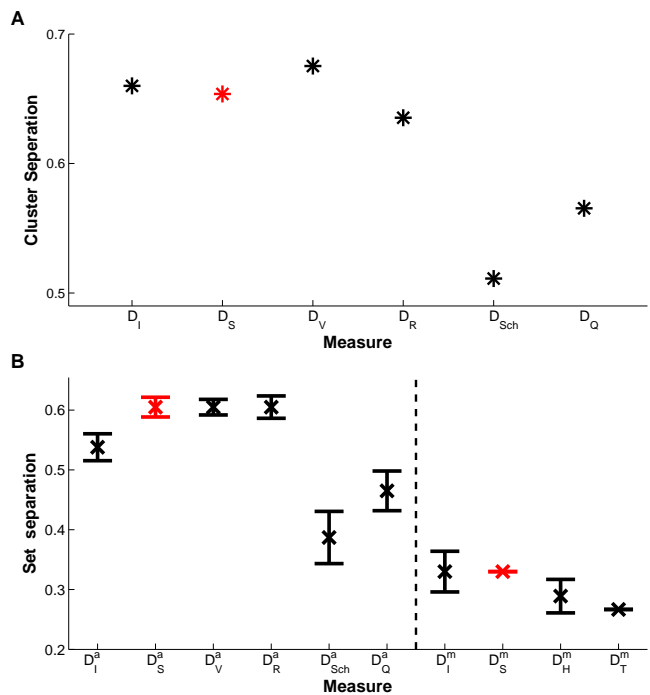}
    \caption{\abb\label{fig:HR-Bi-Multi-Comp} (color online) Controlled comparison of spike train distances on a Hindemarsh-Rose network. A. Comparison of bivariate measures: Separation of clusters.   B. Comparison of multivariate measures: The set separation is defined as the fraction of tests with a Kolmogorov-Smirnov statistic equal to $1$. The dashed line separates the averaged bivariate measures from the multivariate measures. Note that panel A is an extension of Fig. 11 from \citet{Kreuz07c}, and panel B is an extension of Fig. 15 from \citet{Kreuz09a}.}
\end{figure}

Details on the analysis can be found in \citet{Kreuz07c}. In short: We applied the dissimilarity measures to all possible pairs of spike trains and from the resulting pairwise distance matrices we generated hierarchical cluster trees. From these dendrograms we extracted the cluster separation, an indicator which quantifies how well the different measures can separate the three clusters of the network. We here repeat exactly the same analysis for the spike-based distance $D_{\mathrm {S}}$. As shown in Fig. \ref{fig:HR-Bi-Multi-Comp}A both the ISI- and the SPIKE-distance almost match the cluster separation of the best time-scale optimized measure. The poorest cluster separation is obtained for the optimized $D_{\mathrm {Sch}}$.

\subsection{\label{App-ss:HR-Multivariate-Comparison} Multivariate setup: Assessing set separation}

Following \citet{Kreuz09a} we use the same controlled setup with the network of Hindemarsh-Rose neurons to compare the performance of several averaged bivariate and multivariate measures in distinguishing different levels of multi-neuron spike train synchrony. The averaged bivariate measures are the extended variants of the bivariate measures used in Appendix \ref{App-ss:HR-Bivariate-Comparison}, multivariate measures comprise the multivariate ISI- and SPIKE-distances as well as the reliabilities $D_{\mathrm {H}}^{\mathrm {m}}$ by \citet{Hunter98} and $D_{\mathrm {T}}^{\mathrm {m}}$ by \citet{Tiesinga04}. While $D_{\mathrm {H}}^{\mathrm {m}}$ measures the normalized variance of pooled, exponentially convolved spike trains, $D_{\mathrm {T}}^{\mathrm {m}}$ exploits the deviation of pooled spike train statistics from the one obtained for a Poisson process. Again the performance of the time-scale dependent measures is optimized with respect to the time-scale parameter.

In order to gradually change the level of synchrony in a controlled manner, we construct randomly selected sets of spike trains from the two principal clusters of the Hindemarsh-Rose network and vary the set imbalance, the relative contributions of these clusters. For each set imbalance we create five groups of $100$ realizations and then employ the set separation, a simple measure based on Kolmogorov-Smirnov statistics, to quantify how well the distributions of measure values for adjacent set imbalances can be distinguished. We repeat exactly the same analysis for the averaged bivariate SPIKE-distance $D_{\mathrm {S}}^{\mathrm {a}}$ as well as for the multivariate SPIKE-distance $D_{\mathrm {S}}^{\mathrm {m}}$. Results for all measures are shown in Fig. \ref{fig:HR-Bi-Multi-Comp}B.

In this multivariate context the SPIKE-distance $D_{\mathrm {S}}^{\mathrm {a}}$ matches the performance of the best time-scale optimized measures. The ISI-distance $D_{\mathrm {I}}^{\mathrm {a}}$ is very close behind. Overall, the averaged bivariate measures are consistently better at distinguishing different set imbalances than the multivariate measures. Among these the highest set separations are obtained for the ISI-distance $D_{\mathrm {I}}^{\mathrm {m}}$ closely followed by the SPIKE-distance $D_{\mathrm {S}}^{\mathrm {m}}$. This is probably due to the fact that these measures are not invariant to shuffling spikes among the spike trains but rather do take into account the spike train of origin for each individual spike. On the other hand, measures that act on the pooled spike train, such as all measures based on the Peri-Stimulus Time Histogram (PSTH), yield the same value regardless of how spikes are distributed among the different spike trains (cf. also Fig. \ref{fig:SPIKE-PSTH}).

\end{appendix}

\vspace{1cm}

\begin{thanks}
\section{\label{s:Acknowledgement} \textbf{Acknowledgements}}

    We acknowledge useful discussions with S Luccioli, A Politi and A Torcini. We thank J Shlens, GD Field, JL Gauthier, A Sher, MI Grivich, D Petrusca, AM Litke and EJ Chichilnisky for providing the multi-neuron recordings from retinal ganglion cells. We also thank D Gardner and all others involved in the "neurodatabase.org" project, as well as PM Di Lorenzo and JD Victor for sharing their data there. Finally, many thanks to A Morelli for the Hindemarsh-Rose data. TK has been supported by the Marie Curie Individual Outgoing Fellowship "STDP", project No $040576$, DC by grants $2008BE1 00166$ and $2009FI B100087$ of the \textit{Generalitat de Catalunya} and European Social Funds, MG by a Pioneer Postdoctoral Fellowship Award. RGA acknowledges grant $BFU2007 61710$ of the Spanish Ministry of Education and Science.
\end{thanks}


\bibliographystyle{elsart-harv}

\end{document}